\documentclass[aps,twocolumn,showpacs]{revtex4}
\usepackage{epsfig}

\begin{document}


\title{Intrinsic and extrinsic inhomogeneities in mixed-valence
manganites}

\author{B. I. Belevtsev}
\email[]{belevtsev@ilt.kharkov.ua}
\affiliation{B. Verkin Institute for Low Temperature Physics and
Engineering, National Academy of Sciences, pr. Lenina 47, Kharkov
61103, Ukraine}


\begin{abstract}
It is suggested that extrinsic inhomogeneities in mixed-valence
manganites deserve more attention and they should be taking into
account on equal footing with hypothetical phase separation while
examinating experimental data and developing the theoretical
models of influence of stoichiometric and other types of
inhomogeneities on properties of these and other transition-metal
oxides.
\end{abstract}


\maketitle
\par
The structural, magnetic and electron transport properties of
mixed-valence manganites of the type R$_{1-x}$A$_x$MnO$_3$ (where
R is a rare-earth element, A a divalent alkaline-earth element)
attracted much attention of scientific community in the last
decade (see reviews
\cite{ramirez,coey,dagotto,nagaev,kim,dagotto1}). The interest is
caused by observation of huge negative magnetoresistance (MR)
near the Curie temperature, $T_c$, of the
paramagnetic-ferromagnetic transition for manganites with $0.2
\leq x \leq 0.5$. This phenomenon was called ``colossal''
magnetoresistance (CMR) and is expected to be used in advanced
technology. The unique properties of mixed-valence manganites are
determined by complex spin, charge and orbital ordered phases,
and, therefore, are of great fundamental interest for physics of
strongly correlated electrons. At present time, it is believed
that one of the key feature of manganites is their intrinsic
inhomogeneities in the form of coexisting competing ferromagnetic
and antiferomagnetic/paramagnetic phases
\cite{dagotto,nagaev,dagotto1}. This phenomenon is generally
called ``phase separation''. In Refs. \cite{dagotto,dagotto1},
theoretical computational models were developed for two cases: (1)
Electronic phase separation which implies nanocluster
coexistence. (2) Disorder-driven phase separation which leads to
rather large (micrometer size) coexisting clusters. Existence of
the nanoscale as well as micrometer size inhomogeneities in
manganites were corroborated experimentally (see Refs.
\cite{dagotto,nagaev,dagotto1} and references therein). Some
other examples of the phase-separation models can be found in
Refs. \cite{nagaev,mathur,sboychak,iad,yukalov} (actually, there
is a vast literature on the subject, but it can not be cited more
fully in this short communication). On the basis of this, it is
hoped to explain transport and magnetoresistive properties
(including CMR) of manganites taking into account the phase
separation effects.
\par
In spite of enormous theoretical and experimental activity in the
area of the phase separation in manganites, many questions
(sometimes rather simple and naive) remain open. Intrinsic
inhomogeneities are believed to arise for thermodynamical reasons
so that relative fraction of competing phases should depend on
temperature, pressure, and magnetic field. The known experimental
studies give numerous (predominantly indirect though) evidences
of structural and magnetic inhomogeneities in manganites, but are
they in all cases intrinsic? The point is that in all manganites
the extrinsic inhomogeneities are inevitably present (even in
single crystal samples). Extrinsic inhomogeneities arise due to
various technological factors in the sample preparation. They can
cause chemical-composition inhomogeneity (first of all in the
oxygen content), structural inhomogeneities (polycrystalline or
even granular structure), strain inhomogeneities and so on. It is
easy to find in literature a lot of experimental studies in which
finding of phase separation effects is proclaimed, but the
interpretations are often doubtful.  In such cases the effects of
technological inhomogeneities are quite obvious or, at least, can
not be excluded. In some cases the magnetic inhomogeneities,
induced by extrinsic reasons, can depend significantly on
temperature, pressure and magnetic field as well, and their
apparent influence on magnetic and transport properties of
mixed-valence manganites may agree generally with that of
predicted by some of the numerous phase-separation theoretical
models. It should be noted, however, that a quantitative
comparison of the known models with experiment is practically
impossible (or is too ambiguous).
\par
Consider shortly the main sources of extrinsic inhomogeneities.
Mixed-valence manganites are complex perovskite-like oxides
consisting of, at least, four elements. Their properties are very
sensitive to crystal imperfections, especially to the structural,
composition  and other types of inhomogeneity in crystal lattice.
The crystal perfection (and corresponding level of inhomogeneity)
depends strongly on method of preparation, and on preparation
conditions for the given method. In rough outline, the following
methods of manganite growth are used: 1) thin film growth (mostly
with pulsed-laser deposition method); 2) solid-state reaction
method; 3) floating zone method.
\par
Thin manganite films can be prepared highly oriented or even
single-crystal epitaxial with fairly perfect crystal lattice. The
highest values of magnetoresistance were observed in thin films.
But it should be taken into account that films are always in an
inhomogeneous strained state due to inevitable substrate-film
lattice interaction, that induces, as a rule, a considerable
magnetic and magnetoresistance anisotropy \cite{belev}. Due to
strained state, some other film properties (among other things,
the value of $T_c$) can be quite different from these of bulk
materials.
\par
Consider some examples of extrinsic inhomogeneities in films. A
comprehensive and thorough study (with high-resolution electron
microscopy) \cite{lebedev} of epitaxial La$_{1-x}$Ca$_x$MnO$_3$
($x\approx 0.3-0.35$) films grown on SrTiO$_3$ substrate has
revealed that close to substrate a perfectly coherent strained
layer is formed, above which crystal blocks with columnar
structure grow; these blocks and boundary regions between them
accomodate the lattice mismatch between substrate and film.
Boundary regions between the blocks (domains) are
non-stoichiometric, having deficiency of oxygen and of lanthanum.
Similar results are found in Ref. \cite{lu}, where secondary-phase
non-stoichiometric rods were found in La$_{0.7}$Ca$_{0.3}$MnO$_3$
films grown on LaAlO$_3$ and SrTiO$_3$ substrates. The films have
a domain structure, in which the rods are believed to be
responsible for relieving stress during film growth. Magnetic
force microscopy study of pulsed laser deposited
La$_{1-x}$Sr$_x$MnO$_3$ ($x=$0.23 and 0.3) films \cite{soh} have
revealed local FM regions at temperature above the $T_c$ of the
film. These regions with higher $T_c$ were found around the grain
boundaries and attributed to the local variation of strain in
film. The above examples show that even epitaxial films, prepared
at optimal conditions, have inhomogeneous strains and a local
non-stoichiometry, that can play a significant role in transport
and magnetoresistive properties of thin films.
\par
The solid-state reaction (SSR) technique enables preparing ceramic
or polycrystalline samples. The crystal quality (and, therefore,
resistive, magnetoresistive and magnetic properties) of the SSR
samples depends in crucial way on preparation conditions,
especially on sintering and annealing temperature. In samples
prepared with optimal sintering temperature, a fairly sharp
resistive and magnetic transitions near $T_c$ are observed;
whereas, quite different resistive and magnetization behavior is
seen for samples with the same nominal composition, but prepared
at low temperature \cite{joy}. This is to be attributed to
composition and structure inhomogeneity of samples sintered at
low temperatures. In all preparation conditions, however, SSR
samples are always polycrystalline and contain inevitably at least
one source of inhomogeneity: grain boundaries regions. These are
regions of structural, magnetic and stoichiometric disorder, and,
therefore, they have different conducting and magnetic properties
as compared with these inside the grains. Beside this, a rather
appreciable composition inhomogeneities (not associated with
grain boundaries) can not be excluded in SSR samples even when
they are prepared at optimal conditions. The common methods of
checking of stoichiometric inhomogeneity and mixed-phase state
(x-ray powder diffraction or electron microprobe analysis) have
too low accuracy to come to unambiguous conclusion about
composition homogeneity. For example, if a sample is a mixture of
two phases of R$_{1-x}$A$_x$MnO$_3$, composed from the same
elements, but with appreciably different values of $x$ or oxygen
concentration, it is hard or even impossible to see clearly
enough the two-phase state in diffraction pattern, even if volume
fractions of the phases are comparable; whereas, magnetic and
other properties of these phases can be significantly different.
Only non-perovskite-type impurities can be detected quite clearly
down to 2\%. Electronic microprobe elemental analysis has an
accuracy about $\pm 5$~\%, in most cases, leaving room for
stoichiometic disorder within these limits. More powerful, but
much more expensive methods, like neutron diffraction or
small-angle neutron scattering, are not in common use, but even
these methods have their limits of accuracy. Since properties of
manganites are very sensitive to chemical composition and,
therefore, to stoichiometric disorder, no wonder to find in
literature quite different properties of manganites of the same
nominal composition, prepared by SSR method. In spite of the
unavoidable technological inhomogeneity, the SSR method is in
common use for preparation of mixed-valence manganites of various
composition. The reason is that SSR method appears to be not very
sophisticated (at least, at first glance) and does not require an
expensive equipment. At proper experience and rather hard work,
it is possible to obtain polycrystalline samples of rather good
quality with sharp resistive and magnetic transitions. For
example, a generally recognized phase diagram for system
La$_{1-x}$Ca$_x$MnO$_3$ was obtained for SSR polycrystalline
specimens \cite{kim}.
\par
It is easy to find in literature hundreds of papers devoted to
the film or bulk ceramic manganites, but far less studies concern
single-crystal samples. The obvious reason is that it is not so
easy to prepare manganite single crystals. But even
single-crystals prepared by the floating zone method are not free
from defects and extrinsic inhomogeneities. Really, they have
mosaic blocks, twins, inhomogeneous strains, and stoichiometric
disorder \cite{muk1,muk2,aken}.
\par
The experimental data provide, therefore, that the technological
inhomogeneities are unavoidable for any preparation method, and
actually they can be called ``intrinsic'' as well. For this
reason, (i) in many cases it is better to speak about multiphase
coexistence instead of the phase separation; (2) the technological
inhomogeneities should be directly taken into account in new
theoretical models. The latter demand is conditioned by the
circumstance that manganite materials which can be used in an
advanced technology will surely have some crystal imperfections or
inhomogeneities. Moreover, in some cases specific types of
inhomogeneities should be even induced specially to provide
necessary properties. For example, grain boundaries or specially
prepared percolation structures can ensure high MR in low fields
in temperature range far below $T_c$, that may be necessary for
some applications.
\par
As for the phase separation, this concept becomes now, on the one
hand, a commonplace, but, on the other hand, the term is too
general to imply something specific. At interpretation of their
results, experimentalists often speak quite generally about phase
separation or just mention it, meaning not much at that. And how
they can, if at least a dozen of diverse models (suggesting quite
different mechanisms of phase separation) are developed, which,
however, practically can not be numerically compared with
experiment? In spite of this, the phase-separation concept
appears to be very attractive since it can give a quite natural
qualitative explanation for both the huge drop in resistance and
the CMR in vicinity of magnetic transitions in manganites, taking
into account a percolating character of these transitions
\cite{nagaev,sboychak}. Consider, for example,
La$_{1-x}$Ca$_x$MnO$_3$ system. According to Refs.
\cite{mira,shin,zhang,hong,gordon,adams},
paramagnetic-ferromagnetic (PM-FM) transition in this compound is
of first order for the range $0.25 < x < 0.4$. It is found in
these compounds that FM metallic clusters are present well above
$T_c$, while some PM insulating clusters can persist down to a
range far below $T_c$ \cite{teresa,dho,papa}. That seems
naturally for first-order transition where nucleation of the FM
clusters above $T_c$ is quite expected, as well as the presence
of some amount of PM clusters below $T_c$. After all, transition
of this type is hysteretic and depends on the rate of heating or
cooling. In this case a real phase separation and percolation
processes can be expected around $T_c$. Since the PM phase is
insulating and the FM one is metallic, some kind of
insulator-metal transition takes place near $T_c$. The
technological inhomogeneities broaden the temperature range of
the PM-FM transition so that it may appear more smooth and
continuous, like second order transition.
\par
For Ca concentration outside of the above-indicated range, $0.25
< x < 0.4$, the PM-FM transition is found to be of second order
in La$_{1-x}$Ca$_x$MnO$_3$ samples with $x=$ 0.20, 0.40 and 0.45
\cite{hong,rhyne,kim2}. According to the phase diagram for this
system \cite{ramirez,dagotto,kim,dagotto1}, these concentrations
are close to critical ones: $x\approx 0.2$ (which is a border
between the FM metallic and insulating states) and $x=0.5$ (which
is a border between FM metallic and insulating charge-ordered
states). It is clear that unavoidable technological
stoichiometric disorder will have a greater impact on magnetic
transition for samples having nominal Ca concentrations near the
above-mentioned critical values. The $T_c$ value depends rather
strongly on $x$ near these threshold concentrations; whereas, the
concentration dependence of $T_c$ near the optimal doping
($x\approx 0.35$) is rather weak (see the phase diagram in Refs.
\cite{ramirez,dagotto,kim,dagotto1}). In this case, the magnetic
transition for a sample with non-optimal concentration should be
broader, than that for the optimal-doped samples, even if the
level of composition-inhomogeneity is equal in both cases. It can
not be excluded, therefore, that a second order transition found
for these La$_{1-x}$Ca$_x$MnO$_3$ samples is just rather
broadened (smeared) first order transition.
\par
It should be noted that PM-FM transition is found to be of second
order in Sr-doped La$_{1-x}$Sr$_x$MnO$_3$ samples ($x=$0.3 and
0.33) as well \cite{mira,novak}. The Sr manganites are more
conductive than Ca manganites and have much higher $T_c$ (maximum
$T_c$ are about 260~K and 370 K for Ca and Sr manganites,
respectively). It seems that manganites with higher conductivity
and $T_c$ are more prone to second order transition than those
with low conductivity and $T_c$. In homogeneous samples with
perfect crystal lattice the second order transition from PM to FM
state should proceed at once in all sample volume as soon as the
temperature cross $T_c$ going from above. No nuclei of FM phase
above $T_c$, no supercooling or hysteresis phenomena should occur
at this transition. Only thermodynamical fluctuations of the
order parameter (the magnetization) are expected, which, however,
should be confined to narrow critical region around $T_c$
\cite{vonsovsky,belov}. These fluctuations of magnetic order have
usually a rather noticeable effect on ``non-magnetic''
properties, like temperature coefficient of the resistivity, heat
capacity, magnetoresistance, thermal expansion, in vicinity of
$T_c$ \cite{vonsovsky,belov}.
\par
Stoichiometric disorder and non-homogeneous strains of crystal
lattice, which are unavoidable in real manganites due to the
above-indicated technological reasons, can undoubtedly have a
pronounced effect on the second-order PM-FM transitions. This
effect has long been known and considered for simple FM metals
\cite{belov}. Take, for example, as in Ref. \cite{belov}, a
system consisting of multiple phases with different $T_c$. There
is some volume distribution of regions with different $T_c$
within the sample. Availability of interphase transition regions
between different phases should be taken into account as well.
The temperature dependence of the magnetization for this sample
will show somewhat broaden PM-FM transition \cite{belov} (the
temperature width of the transition depends on how wide is the
distribution of $T_c$ in the sample). From that an averaged $T_c$
value can be determined. But some parts of the sample have $T_c$
greater or less than this averaged value. Therefore, it can be
found with some experimental methods that some FM clusters exist
above $T_c$, with their volume increasing when going to $T_c$
from above; whereas, PM clusters can be found below $T_c$, with
their volume fraction decreasing when going down away from $T_c$.
The reason for this behavior is quite obvious taking into account
the sample inhomogeneity. Now, even if every single phase of this
multiphase system undergoes a second order transition, the total
character of transition will not that for homogeneous system. It
will be of percolative nature. If PM and FM phase states differ
drastically in their conductivity, the CMR can be found. Imagine
that size of inhomogeneities is rather small, say, a few
nanometers (which is quite possible for technological
inhomogeneities). Is it possible in this case to attribute the
magnetotransport behavior of this system near the PM-FM
transition to the phase separation effect with some certainty? The
negative answer is obvious since technological inhomogeneities
alone can provide this behavior.
\par
Due to enormous theoretical activity in this area, it is rather
appropriate to believe that phase separation really takes place in
manganites and in other transition-metal oxides (although it is
difficult to make a right choice from numerous propositions of
the phase separation mechanisms). But how to distinguish surely
enough these thermodynamic effects from those of extrinsic
inhomogeneity? It is a really difficult problem. I think that
theoreticians should not disregard the influence of extrinsic
inhomogeneities, but, on the contrary, they should take them into
account in their models quite directly along with intrinsic
inhomogeneities. This necessity was indicated quite clearly in
the paper of Yukalov \cite{yukalov}. One of the principal ideas
of this  paper is that real systems are never free from external
perturbations, that makes the system stochastically unstable even
if external perturbations are infinitesimally small. After all,
extrinsic inhomogeneities can even stimulate appearance of
thermodynamic phase separation, so that some kind of interaction
between them is possible.
\par
In conclusion, at consideration of experimental data for
mixed-valence manganites and developing of theoretical models for
them, the unavoidable influence of extrinsic disorder and
inhomogeneities should always be taken into account. These
inhomogeneities can act separately as well as together with the
suggested intrinsic inhomogeneities (phase separation) and
determine to a great extent the magnetic and magnetotransport
properties of these compounds. Although, for the most part, the
known properties of La$_{1-x}$Ca$_x$MnO$_3$ system near the PM-FM
transition were used here for backing of the above-mentioned point
of view, the general conclusion of this paper is applicable (in
author opinion) to other magnetic transitions in manganites (for
example, for transitions to charge-ordered states) and to related
magnetic transition-metal oxides, like cobaltites
La$_{1-x}$Sr$_x$CoO$_3$.
\par
Author sincerely acknowledges a very useful discussion of some
questions, touched in this note, with Dr. P. A. Joy from National
Chemical Laboratory, Pune, India.



\end{document}